\documentclass[12pt,dvipsnames]{article}
 
\usepackage{iftex}
\ifLuaTeX
  \usepackage{polyglossia} % Default LuaLaTeX
  \usepackage{fontspec} % Default LuaLaTeX
  \setmainlanguage{english}
  \usepackage[nosort]{cite}
\else % LaTeX
  \usepackage[utf8]{inputenc}
  \usepackage[T1]{fontenc}
  \usepackage[nosort]{cite}
  \usepackage[english]{babel} % foreign languages
\fi

\usepackage[top=2cm,bottom=2cm,outer=2.5cm,inner=2.5cm,marginparwidth=2.5cm]{geometry}
\usepackage{fvextra} % patches brackets and symbols
\usepackage{authblk} % Affiliations for authors

\ifLuaTeX
  \usepackage[unicode]{hyperref}
  \usepackage{pdftexcmds} % define pdfstrcmp in LuaLaTeX
  \makeatletter
  \let\pdfstrcmp\pdf@strcmp
  \makeatother
\else
  \usepackage{hyperref}
\fi

\usepackage{amsmath}
\usepackage{amsfonts}
\usepackage{amssymb}
\usepackage{tensor} % Tensor indices
\usepackage{braket} % Bra-Ket dirac notation
\usepackage{empheq} % Brackets on block of equations
\usepackage{romanbar} % Roman numerals
\usepackage{dsfont} % identity matrix symbol
\usepackage{stmaryrd} % super-Lie brackets

\usepackage{lmodern} % Computer modern font
\usepackage{caption}
\usepackage{subcaption} % Subfigures
\usepackage{titlesec} % Custom sections, titles, etc.
\usepackage{fancyhdr} % Header
\usepackage{abstract} % Include abstracts
\usepackage[super]{nth}
\usepackage[colorinlistoftodos,prependcaption,textsize=tiny]{todonotes} 
\usepackage{tocloft} % Layout of table of contents

\usepackage{graphicx} % Images
\usepackage{multirow} % Fancy tables
\usepackage{array} % For new array types
\usepackage{tikz}
\usepackage[compat=1.1.0]{tikz-feynman} % Feynman diagrams
\usepackage{pgf}
\usepackage{pgfplots} % Fancy plots
\usepackage{dynkin-diagrams}
\usepackage{hhline}

\usetikzlibrary{arrows,shapes,calc,intersections,fadings,positioning,shadows.blur,decorations.pathreplacing,external,backgrounds}
\usepackage{currfile} % Filename for exports

\immediate\write18{mkdir -p TikZ}     %% Create `pgf-img` directory
\tikzexternaldisable
\newcommand*\setmyname{%\
  \expandafter\tikzsetfigurename\expandafter{\currfilebase-}%
}
\AtBeginEnvironment{tikzpicture}{\setmyname}

\hypersetup{
pdfstartview={FitH},
pdftitle={Observations on BPS observables in 6d},
pdfauthor={N. Drukker, M. Trépanier},
pdfsubject={},
pdfcreator={},
pdfkeywords={},
colorlinks=true,
citecolor=MidnightBlue,
linkcolor=MidnightBlue,
urlcolor=MidnightBlue
}

\numberwithin{equation}{section}
\newcommand\frontmatter{%
  \clearpage
  \pagenumbering{roman}
}

\newcommand\mainmatter{%
  \clearpage
  \pagenumbering{arabic}
}

\usepackage{mathtools}

\DeclareMathOperator{\tr}{tr}
\newcommand{\vev}[1]{\left\langle #1 \right\rangle}

\newcommand{\norm}[1]{\left| #1 \right|}

\DeclareMathOperator{\vol}{\mathrm{vol}}

\newcommand{\RicciScalar}{\mathcal{R}}
\newcommand{\cAnomaly}{\left( \partial n \right)^2}

\def\im{i}

\newcommand{\IIFundForm}{\Romanbar{2}}
\newcommand{\diff}{\mathrm{d}}

\def\bC {\mathbb{C}}

\def\bH {\mathbb{H}}

\def\bP {\mathbb{P}}

\def\bR {\mathbb{R}}

\def\aQ{{\mathsf{Q}}}
\def\aS{{\mathsf{S}}}

\def\cN{{\mathcal{N}}}
\def\cK{{\mathcal{K}}}

\newcommand{\su}{\mathfrak{su}}

\newcommand{\osp}{\mathfrak{osp}}
\newcommand{\sof}{\mathfrak{so}}
\newcommand{\spf}{\mathfrak{sp}}

\newcommand{\bea}{\begin{eqnarray}}
\newcommand{\eea}{\end{eqnarray}}
\newcommand{\beq}{\begin{equation}}
\newcommand{\eeq}{\end{equation}}
\newcommand{\bal}{\begin{equation}\begin{aligned}}
\newcommand{\eal}{\end{aligned} \end{equation}}

\def\tC{\hyperlink{tC}{\bf Type-$\bC$}}
\def\tR{\hyperlink{tR}{\bf Type-$\bR$}}
\def\tH{\hyperlink{tH}{\bf Type-$\bH$}}
\def\tL{\hyperlink{tL}{\bf Type-L}}
\def\tS{\hyperlink{tS}{\bf Type-S}}
\def\tN{\hyperlink{tN}{\bf Type-N}}

\title{
Observations on BPS observables in 6d}
\author{Nadav Drukker\thanks{\href{mailto:nadav.drukker@gmail.com}{nadav.drukker@gmail.com}}\ }
\author{Maxime Tr\'epanier\thanks{\href{mailto:maxime.trepanier@kcl.ac.uk}{maxime.trepanier@kcl.ac.uk}}}
\affil{\it Department of Mathematics, King's College London,\\London, WC2R
2LS, United Kingdom}
\date{}

\begin{document}

\frontmatter
\maketitle
\thispagestyle{empty}

\begin{abstract}
We study 
possible geometries and R-symmetry breaking patterns that lead 
to globally BPS surface operators in the six dimensional $\cN=(2,0)$ theory. 
We find four main classes of solutions in different 
subspaces of $\bR^6$ and a multitude of subclasses and specific examples. 
We prove that these constructions lead to supersymmety preserving 
observables and count the number of preserved supercharges. We discuss 
the underlying geometry, calculate their anomalies and present analogous 
structures for the holographic dual M2-branes in $AdS_7\times S^4$. 
We also comment on the dimensional reduction of these observables to line and 
surface operators in 4d and 5d theories. This rich spectrum of operators are 
proposed as the simplest and most natural observables of this mysterious theory.
\end{abstract}

\mainmatter

\tableofcontents

\section{Introduction}

In the quest to ascend the precipice of the 6d $\cN=(2,0)$ theory, we are
in great need of edges or cracks to hold on to. Surface operators are ideal
candidates, being the intrinsic, low dimensional, non-local observables of these
theories~\cite{witten:1995zh,Strominger:1995ac,ganor:1996nf}. Already, recent work combining the powerful techniques of the
superconformal index~\cite{Bullimore:2014upa,Chalabi:2020iie}, holographic
entanglement entropy~\cite{DHoker:2008lup,DHoker:2008rje,Gentle:2015jma,
Rodgers:2018mvq,Jensen:2018rxu,Estes:2018tnu} and defect CFT
techniques~\cite{Drukker:2020swu,Drukker:2020atp, Wang:2020xkc} 
lead to the exact calculation of the anomaly coefficients (see Section~\ref{sec:anomaly}), 
thus giving the leading property of surface operators: the logarithmic divergence 
in their expactation values, which is an analog of the anomalous dimensions 
of local operators \cite{deser:1993yx,Boulanger:2007st,Schwimmer:2008yh,solodukhin:2008dh,Drukker:2020dcz}.

A distinguished class of surface operators are BPS observables, i.e. those
that preserve some of the supercharges of the theory. For these operators,
supersymmetry affords many simplifications and, from the experience with
BPS Wilson loops in lower dimensional theories 
\cite{erickson:2000af, drukker:2000rr, pestun:2007rz, 
Kapustin:2009kz, Drukker:2009hy, Marino:2009jd, 
drukker:2007qr, correa:2012at, Fiol:2012sg}, 
we expect BPS surface operators to be amenable to a variety of 
nonperturbative techniques, providing an ideal playground to
learn more about the mysterious 6d $\cN = (2,0)$ theories.

The obvious examples of BPS surface operators are the plane~\cite{Howe:1997ue,maldacena:1998im}
and the sphere~\cite{berenstein:1998ij}, which preserve half of the supercharges
and are analysed in detail in~\cite{DHoker:2008lup,DHoker:2008rje,Gentle:2015jma,
Rodgers:2018mvq,Jensen:2018rxu,Estes:2018tnu,Drukker:2020atp,chen:2007ir,chen:2008ds,mori:2014tca,Cremonini:2020zyn}.
There are no explicit expressions for the surface operators beyond the abelian 
theory and the holographic theory at large $N$, but they should be labeled by a representation 
of the ADE algebra of the theory. As far as the global symmetries, the 1/2 BPS operator 
is defined by a choice of plane in 6d, which breaks the 6d conformal symmetry
$\sof(2,6)$ to $\sof(2,2) \times \sof(4)$, and a breaking of the $\sof(5)$
R-symmetry to $\sof(4)$, which can be realised by choosing a point on $S^4$.
Accounting for supercharges, the full preserved symmetry is an $\osp(4^*|2)^2$
subalgebra of the $\osp(8^*|4)$ symmetry algebra of the vacuum.

This analysis can be generalised to surface operators
with an arbitrary geometry $\Sigma \to \bR^6$ and a local breaking of
R-symmetry $n : \Sigma \to S^4$~\cite{Drukker:2020dcz}. Pointwise these
operators preserve half of the supersymmetries, but typically the
supersymmetries are not compatible over the surface, thus can be called 
``locally BPS''.

In this paper we present several large families of truly BPS surface operators, 
requiring the supercharges
to be compatible over the full surface.  As a simple illustration, consider two
planes extended respectively along $(x^1,x^2)$ and $(x^1, x^3)$. For the first
we take $n^{I} = \delta^{I1}$, for the second, $n^{I} = \delta^{I2}$. The two
operators so defined share half of their supercharges, and as the planes
intersect along a line, we can take half of each and form a corner (which is
then 1/4 BPS). The operators presented below are massive generalisations of this
example.

A word on notations: we use $x^\mu$ with $\mu = 1, \dots, 6$ as coordinates of 6d 
Euclidean space with metric $\delta_{\mu\nu}$. Spacetime spinor indices are 
$\alpha$ and $\dot\alpha$ respectively for chiral and anti-chiral spinors, and
for the fundamental of $\spf(4)$ we use the indices $A$, all take four values.
For the R-symmetry vectors (fundamental of $\sof(5)$) we employ $I,J =
1,\dots,5$. On the surface $\Sigma$ we use coordinates $\sigma^a=(u,v)$ with $a
= 1,2$ and the metric induced from the embedding $h_{ab} = \partial_a x^\mu
\partial_b x^\mu$ with its inverse $h^{ab}$. We also use $\varepsilon^{ab}$ to denote
the Levi-Civita tensor density (including a factor $\det(h_{ab})^{-1/2}$).

Our construction relies on spinor geometry.  To formulate the problem, recall
that the supercharges of the $\cN=(2,0)$ theory are parametrised by two
16-component constant spinors $\varepsilon_0$ and $\bar\varepsilon_1$. They are
respectively chiral and anti-chiral in 6d and transform also under $\spf(4)$.
Since we work in Euclidean signature our spinors do not satisfy a reality
condition. $\varepsilon_0$ parametrises the 16 super-Poincar\'e symmetries,
$\aQ$, and $\varepsilon_1$ the 16 superconformal ones, $\aS$.  Together
$\varepsilon_0 + \bar{\varepsilon}_1 (x \cdot \bar{\gamma})$ are the conformal
Killing spinors in flat space.

We use $\gamma^\mu$ and $\bar\gamma^\mu$ for the spatial gamma matrices in the chiral basis 
and $\rho^I$ for the $\sof(5)$ gamma matrices. They satisfy
\beq
  \bar{\gamma}_\mu \gamma_\nu + \bar{\gamma}_\nu \gamma_\mu = 2 \delta_{\mu\nu}\,, 
  \qquad
   \gamma_\mu \bar{\gamma}_\nu + \gamma_\nu \bar{\gamma}_\mu = 2 \delta_{\mu\nu}\,, 
   \qquad
    \left\{ \rho_I, \rho_J \right\} = 2 \delta_{IJ}\,,
\eeq
and $\gamma_\mu$ and $\rho_I$ commute. Explicitly the indices are
$(\gamma_\mu)_{\alpha}{}^{\dot{\alpha}}$,
$(\bar{\gamma}_\mu)_{\dot{\alpha}}{}^{\alpha}$, $(\rho_I)_A{}^B$,
$(\varepsilon_0)^{\alpha A}$ and $(\bar{\varepsilon}_1)^{\dot{\alpha} A}$, but
we suppress them as there is no ambiguity.
We also use the standard shorthand notation $\gamma_{\mu\nu} \equiv \gamma_{[\mu}
\bar{\gamma}_{\nu]}$ for antisymmetrised products.

With these notations in hand we can be more precise about the conditions
required to preserve supersymmetries.  At any point along the surface, the
supercharges preserved by the tangent plane to $\Sigma$ and the choice of $n$ in
$S^4$ are those satisfying the projector equation
\beq
\left[ \varepsilon_0 + \bar{\varepsilon}_1 (x^\rho \bar{\gamma}_\rho) \right]
\left[ \frac{\im}{2}\varepsilon^{ab} 
\partial_a x^\mu \partial_b x^\nu\gamma_{\mu\nu}
+ n^I \rho_I \right] = 0\,.
\label{eqn:proj}
\eeq
This is a set of coupled equations depending on the embedding of the surface
$x^\mu(u,v)$ and the scalar coupling along the surface $n^I(u,v)$. If $n$ is a
unit vector, then this matrix is half-rank, so pointwise half the components of
$\varepsilon_0, \bar\varepsilon_1$ satisfy it, which is the  
``locally BPS'' notion already mentioned above.

To find global solutions to these equations with the
same $\varepsilon_0$, $\bar{\varepsilon}_1$ along the surface, 
we do not attempt a full classification but
present four main classes of solutions. These four classes have $\Sigma$  
in different subspaces of $\bR^6$ and/or
satisfying different constraints and with appropriate choices of $n^I$. Each of
the four classes has many examples with extra properties. To avoid clutter, we
give special names to only two of those subclasses, and all the other special
examples are explained in the appropriate sections. The main examples are:

\paragraph[]{\hypertarget{tR}{\bf Type-$\bR$:}}
Such surfaces are the product of a curve 
$x^I(u) \subset \bR^5$ ($I = 1, \dots, 5$) and the $x^6=v$ direction. To guarantee 
that they are BPS we should choose the scalar coupling to be tangent to the curve
\beq
n^I(u,v) = \frac{\partial_u{x}^I}{|\partial_u x|}\,.
\label{eqn:typeR}
\eeq
The notation $x^I$ instead of $x^\mu$ is common in our examples, as in topological 
twisting, and is a crucial ingredient to get BPS observables.
Examples of surfaces in this class were previously studied in~\cite{Lee:2006gqa}.

This choice of identification allows for an immediate generalisation 
of multiplying the right hand side by a fixed $SO(5)$ 
matrix (or indeed replace $x^6$ by any other line), so in total $S^5\times SO(5)$ of examples. 
Such modifications are also possible to all the 
classes below, but we do not worry about counting these generalisations and fix 
convenient representatives.

\paragraph[]{\hypertarget{tC}{\bf Type-$\bC$:}}
Viewing $\bR^6=\bC^3$ (with any standard complex structure), we can take the surface 
to be any holomorphic curve with a fixed scalar direction, say $n^1$.

\paragraph[]{\hypertarget{tH}{\bf Type-$\bH$:}}
Restricting the surfaces to $\bR^4$, we can take an arbitrary oriented surface and get 
a BPS observable, as long as 
\beq
n^I = \frac{1}{2} \eta^I_{\mu\nu} \partial_a x^\mu \partial_b x^\nu\varepsilon^{ab},
\label{eqn:typeH}
\eeq
where $\eta$ is the 't~Hooft chiral symbol (see \eqref{eqn:eta}), and $I= 1,2,3$, 
so $n^I$ are in $S^2$.
This has a nice interpretation in terms of the Gauss map, which we explain in
Section~\ref{sec:geom}.  Particular special subclasses are surfaces with 
fixed $n^I=(0,0,1)$ which are holomorphic with respect to an appropriate 
choice of complex structure, so 
they overlap with \tC\ (except for the exchange of $n^1\to n^3$). 
Another special subclass are Lagrangian submanifolds, for which 
$n^3=0$. We denote those as \hypertarget{tL}{{\bf Type-L}}. Lastly we should 
mention that for conical surfaces, the conformal transformation to surfaces 
in $\bR\times S^5$ have been described and studied in great detail 
in \cite{mezei:2018url}.

\paragraph[]{\hypertarget{tS}{\bf Type-S:}}
Any surfaces within an $S^3\subset\bR^6$ is BPS provided we choose
\beq
n^I = \frac{1}{2} \tensor{\varepsilon}{^{IJKL}}\varepsilon^{ab} \partial_a x^J \partial_b x^K x^L
\,.
\label{eqn:typeS}
\eeq
$x^L$ appears explicitly and this expression assumes a sphere centred around
the origin. We also assume a unit sphere, our results can be trivially extended
to a sphere of arbitrary radius by dimensional analysis.
These surfaces are different from the restriction of the \tH\ ansatz to $S^3\subset\bR^4$, 
but they do overlap at the vicinity of the north pole, where $x^\mu\sim
\delta^{\mu4}$ and using the 3-index antisymmetric tensor, both
\eqref{eqn:typeH} and \eqref{eqn:typeS} reduce at leading order in $\norm{x^\mu - \delta^{\mu 4}}$ to
\beq
n^I = \frac{1}{2} \tensor{\varepsilon}{^{IJK}} \partial_a x^J \partial_b x^K 
\varepsilon^{ab}\,.
\label{eqn:typeN}
\eeq
We will refer to this subclass as \hypertarget{tN}{{\bf Type-N}}.

The paper is organised by themes. In Section~\ref{sec:geom} we present the
geometry underlying these constructions. The proof of supersymmetry for these
classes of examples is contained in Section~\ref{sec:susy}. We discuss the
Weyl anomaly in Section~\ref{sec:anomaly}, the reduction to line and surface
operators in lower dimensional theories in Section~\ref{sec:reduction},
and the holographic duals in Section~\ref{sec:holography}. The properties
pertaining to each of the constructions above are scattered between these
sections. To aid the reader we provided extensive cross referencing. 
Also, we summarize the main classification in Table~\ref{tab:types} and some of
the most important examples in Table~\ref{tab:examples}, with links to the relevant
pages where they are discussed.

\renewcommand{\arraystretch}{1.15}
\begin{table}[h!]
  \begin{center}
    \begin{tabular}{|c|c|c|c|c|c|c|}
      \hline
      \textbf{Type} &\bf Geometry &\bf $n^I$ in &\bf SUSYs & \bf Also in & \bf
      Anomaly &\bf 3-form \\\hhline{|=|=|=|=|=|=|=|}
      \multirow{4}{*}{\hyperlink{tR}{$\bR$}}
      & $\bR\times\gamma$, $\gamma\subset\bR^5$ & \hyperref[eqn:typeR]{$S^4$} &
      \hyperlink{tRsusy}{1 $\aQ$} && \multirow{4}{*}{0}&
      \multirow{4}{*}{\eqref{eqn:typeR3form}} \\
      & $\bR\times\gamma$, $\gamma\subset\bR^4$ & \hyperref[eqn:typeR]{$S^3$} &
      \hyperlink{tRsusy}{1 $\aQ$} &&& \\
      & $\bR\times\gamma$, $\gamma\subset\bR^3$ & \hyperref[eqn:typeR]{$S^2$} &
      \hyperlink{tRsusy}{2 $\aQ$} &$\bH$&& \\
      & $\bR\times\gamma$, $\gamma\subset\bR^2$ & \hyperref[eqn:typeR]{$S^1$} &
      \hyperlink{tRsusy}{4 $\aQ$} &$\bH$&& \\
      \hline
      \multirow{2}{*}{\hyperlink{tC}{$\bC$}}
      & $\Sigma\subset\bC^3$ (holo.) & point &
      \hyperlink{tCsusy}{2 $\aQ$} && \multirow{2}{*}{\eqref{eqn:typeCanomaly}} & 
      \multirow{2}{*}{\eqref{eqn:typeC3form}}\\
      & $\Sigma\subset\bC^2$ (holo.) & point &
      \hyperlink{tCsusy}{4 $\aQ$} & $\bH$, L && \\
      \hline
      \hyperlink{tH}{$\bH$} & $\Sigma\subset\bR^4$ & \hyperref[eqn:typeH]{$S^2$}&
      \hyperlink{tHsusy}{1 $\aQ$} && \multirow{3}{*}{\eqref{eqn:typeHanomaly}} & \multirow{3}{*}{\eqref{eqn:typeH3form}} \\
      \multirow{2}{*}{\footnotesize subclass $\begin{dcases*}
        \hyperlink{tL}{L} \\ \hyperlink{tN}{N} \\ \end{dcases*}$} &
      \hyperlink{lagdef}{Lagrangian} & $S^1$ &
      \hyperlink{tHsusy}{2 $\aQ$} && & \\
      & $\Sigma \subset \bR^3$ & \hyperref[eqn:typeN]{$S^2$} &
      \hyperlink{tHsusy}{2 $\aQ$} & (S) && \\
      \hline
      \hyperlink{tS}{S} & $\Sigma \subset S^3$ & \hyperref[eqn:typeS]{$S^3$} &
      \hyperlink{tSsusy}{2 $(\aQ+\aS)$} && \eqref{eqn:typeSanomaly} & 
      \eqref{eqn:phiScomplex}\\
      \hline
    \end{tabular}
    \caption{Main classes and subclasses of BPS surface operators. For each
      geometry we list the minimal number of preserved supercharges, which type
      it belongs to, and list the equation with its anomaly and the associated 3-form 
      defined on a subspace of $AdS_7 \times S^4$ that is compatible with supersymmetry.
      \label{tab:types} }
  \end{center}
\end{table}

\begin{table}[h!]
  \begin{center}
    \label{tab:table2}
    \begin{tabular}{|c|c|c|c|c|c|}
      \hline
      \textbf{Name}& \bf Type &\bf Geometry &\bf $n^I$ in &\bf SUSYs&\bf see\\
      \hhline{|=|=|=|=|=|=|}
      \multirow{2}{*}{cones} &$\bH$& over $\gamma\subset S^3$ & $S^2$ &
      $\aQ, \aS$ & \multirow{2}{*}{\!\!\cite{mezei:2018url}}\\
      &$\bH$, N& over $\gamma\subset S^2$ & $S^2$ &
      $2 \aQ, 2\aS$&\\
      crease &$\bR$, $\bH$, N& 2 half-planes & $S^0$ &
      $4 \aQ, 4\aS \subset \osp(4^*|2)$& \cite{Agmon:2020pde} \\
      \hline
      \multirow{2}{*}{tori}&S& $T^2\subset S^3$ & $S^4$ & $2(\aQ+\aS)$&\\
      &$\bH$, L & $T^2\subset \bR^4$ & $S^1$ &
      $2\aQ, 2\aS$ &\\
      \hline
      \multirow{3}{*}{spheres} &$\bH$, N& $S^2\subset \bR^3$ & $S^2$ &
      $2 \aQ, 2\aS$&\\
      &S& latitude $S^2\subset S^3$ & $S^3$ & $4 (\aQ+\aS)$ &\\
      &S& large $S^2\subset S^3$ & point &
      $ 16 (\aQ +\aS) \subset \osp(4^*|2)^2$ &\cite{berenstein:1998ij}\\
      \hline
      plane &$\bR$ $\bC$ $\bH$ L N& $\bR^2$ & point & $8 \aQ, 8\aS \subset
      \osp(4^*|2)^2$ & \cite{maldacena:1998im}\\
      \hline
    \end{tabular}
    \caption[]{Examples of BPS operators with enhanced supersymmetry. \label{tab:examples}}
  \end{center}
\end{table}

\section{Geometry}
\label{sec:geom}

In all our examples, we construct BPS observables by appropriately choosing 
the surface and a vector $n$ over it. The details of these choices rely on 
varied notions of geometry of surfaces, which we review in turn for each type.

\paragraph[]{\tR:}
These surfaces are defined in terms of a curve in $\bR^5$ and $n^I$ \eqref{eqn:typeR} is its 
(normalised) tangent vector. Clearly for a general enough curve we can get any closed path 
$n^I(u)$ on $S^4$. If the curve is restricted to an $\bR^d$ subspace, then 
$n^I$ are constrained to an $S^{d-1}$. As we show in the next section, 
such loops preserve $\max(1,2^{4-d})$ of the Poincar\'e supercharges $\aQ$.

\paragraph[]{\tN:}
To ease into the other examples, let us start with this subclass of surfaces in $\bR^3$ where 
$n^I$ is the (unit) normal vector with a chosen orientation \eqref{eqn:typeN}. 
This map from a surface to $S^2$ is known as the Gauss map and its degree is
half of the Euler characteristic of the surface, which is important for the
calculation of the anomaly below.

\paragraph[]{\tH:}
The Gauss map has an interesting generalisation to surfaces in
$\bR^4$~\cite{hoffman1985gauss}, which we can use to pick our vector $n$. At any
point on the surface the tangent space is a
plane in $\bR^4$. The space of these planes is the Grassmanian $G_2(\bR^4)$,
which is a homogenous space defined as the quotient of the orthogonal group
$SO(n)$
\beq
  G_k(\bR^d) = \frac{SO(d)}{SO(k) \times SO(d-k)}.
  \label{eqn:grassmannian}
\eeq
For planes in 4d, the image of the Gauss map is thus a point on $G_2(\bR^4) =
S^2\times S^2$.

The ansatz \eqref{eqn:typeH} restricts $n^I$ to $S^2$, which is just one of the two spheres 
of the Gauss map. To understand it, recall that 2-forms in 4d can be decomposed
into self-dual and anti-self-dual 2-forms. $\eta^I_{\mu\nu}$ projects to the self-dual
part, which are in the $({\bf 3},{\bf 1})$ representation of $\su(2)_L\times
\su(2)_R\simeq \sof(4)$. Another way to see it is that as a hyperk\"aler
manifold, $\bR^4$ has $S^2$ worth of K\"ahler forms $\omega^I$ (compatible with
the orientation $\varepsilon_{1234}=1$) and any surface is locally holomorphic
with respect to one of them. We can write a basis for them in terms of the
't~Hooft symbols
\beq
  \omega^I = \frac{1}{2} \eta^I_{\mu\nu} \diff x^\mu \wedge \diff x^\nu\,.
  \label{eqn:kahler}
\eeq
Explicitly we use
\bal
\label{eqn:eta}
\omega^3 = \diff x^1 \wedge \diff x^2 + \diff x^3 \wedge \diff x^4\,,\\
\omega^2 = \diff x^3 \wedge \diff x^1 + \diff x^2 \wedge \diff x^4\,,\\
\omega^1 = \diff x^1 \wedge \diff x^4 + \diff x^2 \wedge \diff x^3\,.\\
\eal
If the map degenerates to a point $n^3=1$, 
the self-dual projection of the tangent space is a linear combination of the 
$(x^1,x^2)$ and $(x^3,x^4)$ planes, and in fact 
is holomorphic with respect the complex structure associated to $\omega^3$, hence of \tC. 
The converse to this is a restriction to $n^3 = 0$ everywhere, which means
that the pullback to $\Sigma$ of $\omega_3$ vanishes. Surfaces satisfying this
condition are \hypertarget{lagdef}{Lagrangian submanifold}, as in \tL, and the half of the Gauss map 
represented by $n^I$ reduces to an equator of $S^2$.

\paragraph[]{\tS:}
These are also surfaces in $\bR^4$, but now restricted to $S^3$ with $n^I$ as in \eqref{eqn:typeS}. One 
way to understand this construction is that locally we have a surface in $\bR^3$ with a version of the \tN\ ansatz, 
so a local Gauss map, where the target $S^2$ is the one perpendicular to the vector $x$.

Note that restricting $\Sigma$ to any great $S^2\subset S^3$ makes $n^I$ a constant, so these are 1/2 
BPS spheres, conformal to the plane. A more general subclass are surfaces foliated by 
arcs between the north and south pole, which describe a geometry reminiscent of a
banana. In this case $n^I$ belongs to an $S^2$, which is the sphere we get
for the north pole \tN\ surfaces. These ``banana'' surfaces can be obtained by a conformal
transformation (stereographic projection) of the cones in $\bR^3$ and in fact the 
value of $n^I$ exactly matches those of cones of \tN\ also away from the north pole. 
These surfaces also have \hyperlink{cone}{enhanced supersymmetry}.

One more subclass of \tS\ are non-maximal or ``latitude'' spheres. They resemble in 
some ways the latitude Wilson loops of \cite{Drukker:2006ga, drukker:2007qr} and their $n^I$ image is also 
a latitude sphere there. One can think of those as interpolating between the 1/2 BPS 
sphere and the \tN\ sphere. Similar interpolations are also possible between \tL\ and \tC\ 
within \tH, which are surfaces with fixed $n^3$, so the Gauss map is into a 
latitude on $S^2$.

\paragraph[]{\tC:}
In this case we choose a complex structure on $\bR^6$ and take any holomorphic curve. 
Regardless of how complicated the surface is, a constant $n^I=\delta^{I1}$ yields a BPS 
surface. For a generic surface, as we show in the next section, it preserves two
$\aQ$'s. If it is located in a $\bC^2$ subspace (so overlapping with \tH) it preserves
four, and the plane $\bC$ preserves eight.

\section{Supersymmetry}
\label{sec:susy}

The approach for proving that our examples preserve some supersymmetries, or in
fact for constructing the examples relies on a generalisation of the idea
already presented in the introduction. For each class of examples we can reduce
the projector equation~\eqref{eqn:proj} over the whole surface to a set of independent
constraints on $\varepsilon_0$ and $\bar{\varepsilon}_1$ associated with each 
independent tangent plane of the surface (and its vector $n$).

The number of supercharges preserved by a given surface operator depends on the
number of constraints needed to satisfy~\eqref{eqn:proj}. Since each
independent constraint is half-rank (this can be seen pointwise
from the projector equation) and $\varepsilon$ has 32 components, in the most
general case we could impose five independent conditions leaving a single
Killing spinor. Each Killing spinor $\varepsilon$ satisfying the constraints
parametrises a supersymmetry transformation given by
\begin{align}
  \varepsilon_0 \aQ + \bar{\varepsilon}_1 \aS\,.
\end{align}

The analysis of supersymmetry captures many geometrical properties of the
surfaces and the associated map $n^I$ presented in Section~\ref{sec:geom}.

\paragraph[]{\hypertarget{tRsusy}{\tR}:}
This case is particularly simple and closely resembles the beautiful
construction of~\cite{zarembo:2002an}.
Choosing matching bases for the 2 unit $\sof(5)$ vectors $\partial_u x^I$ and $n^I$ and
identifying them via~\eqref{eqn:typeR}, the projector \eqref{eqn:proj} factorises as
\begin{align}
  \left[ \varepsilon_0 + \bar{\varepsilon}_1 (x^\rho \bar{\gamma}_\rho) \right]
  ( i \gamma_{I6} + \rho_I )\,n^I(u)\,.
\end{align}

For a generic curve $x^I(u)$, all special supersymmetries are broken
($\bar\varepsilon_1 = 0$) and we focus only on $\varepsilon_0$.  At any point,
the constraint imposed on the Killing spinor is a linear combination of the five
projectors acting on $\varepsilon_0$
\beq
\label{eqn:typeRproj}
i \gamma_{I6} + \rho_I\,,\qquad I=1,\cdots,5\,.
\eeq
To count the number of independent constraint we multiply them
from the right by $\frac{i}{2}\gamma_{56}$ and define
\beq
a^\dagger_I=\frac{1}{2}(\gamma_{I5} + i\gamma_{56}\rho_I)\,,
\qquad
a_I=\frac{1}{2}(-\gamma_{I5} + i\gamma_{56}\rho_I)\,,
\qquad I=1,\cdots,4\,.
\eeq
These operators satisfy an oscillator algebra, which shows they are independent
\beq
  \label{eqn:ladderalgebra}
 \{ a_I, a^\dagger_J\} = \delta_{IJ}\,,
  \qquad
  \{ a_I, a_J\} =\{ a_I^\dagger, a_J^\dagger\} = 0\,.
\eeq
Note that since our gamma matrices are chiral, they satisfy
$\gamma_1 \bar{\gamma}_2 \gamma_3 \bar\gamma_4 \gamma_5 \bar{\gamma}_6= i$, or
equivalently $\gamma_1 \bar\gamma_2 \gamma_3 \bar\gamma_4 = -i\gamma_{56}$. Likewise, $\rho_1\rho_2\rho_3\rho_4=\rho_5$, so the spinor
annihilated by all four conditions $a_I^\dagger$ also satisfies $\varepsilon_0(i
\gamma_{I5} + \rho_5)=0$.

Hence if the curve is a straight line, $n$ is a constant vector and the
projector equation imposes a single condition. This is the plane, which
preserves eight $\aQ$ supercharges (and exceptionally also eight $\aS$). For each
extra dimension that the curve visits, the number of supercharges reduces by a
half, except for the last dimension, that does not involve a new projector.
Generically there are no solutions with nonzero $\varepsilon_1$, so no preserved
special supersymmetries $\aS$.

\paragraph[]{\hypertarget{tCsusy}{\tC}:}
To make our discussion concrete it is convenient to choose an explicit complex
structure for $\bR^6$, represented here via the K\"ahler form
\beq
  \frac{1}{2} \omega_{\mu\nu} \diff x^\mu \diff x^\nu
  = \diff x^1 \wedge \diff x^2 + \diff x^3 \wedge \diff x^4 + \diff x^5 \wedge
  \diff x^6\,.
  \label{eqn:kahlerformr6}
\eeq
Holomorphic curves with respect to this complex structure are those satisfying
the Cauchy-Riemann equations
\beq
  \partial_a x^\mu =
  \omega^\mu{}_\nu \varepsilon_{ab} h^{bc} \partial_c x^\nu\,.
  \label{eqn:holomorphic}
\eeq
The natural constraints we impose on $\varepsilon_0$ are then
\beq
\varepsilon_0(i\gamma_\mu+\omega_\mu{}^\nu\gamma_\nu\rho_1)=0\,.
\label{eqn:typeCproj}
\eeq
Of those six equations only the ones with $\mu=1,3,5$ are independent, as for example
\beq
\varepsilon_0(i\gamma_2+\omega_2{}^\nu\gamma_\nu\rho_1)
=\varepsilon_0(i\gamma_2-\gamma_1\rho_1)
=\varepsilon_0(\gamma_2\rho_1+i\gamma_1)i\rho_1
=\varepsilon_0(i\gamma_1+\omega_1{}^\nu\gamma_\nu\rho_1)i\rho_1
=0\,.
\eeq
Now for a generic holomorphic curve, we use the above conditions on 
$\varepsilon_0$ and then the Cauchy-Riemann equation to write
\beq
\varepsilon_0\,i\gamma_{\mu\nu}
\partial_a x^\mu \partial_b x^\nu \varepsilon^{ab}
=-\varepsilon_0\,\omega_{\mu}{}^\sigma\gamma_\sigma\bar\gamma_{\nu}\rho_1
\partial_a x^\mu \partial_b x^\nu \varepsilon^{ab}
=-\varepsilon_0\gamma_\sigma\bar\gamma_{\nu}\rho_1
\partial_a x^\sigma \partial_b x^\nu h^{ab}\,.
\eeq
The symmetry of the term on the right under $\sigma\leftrightarrow\nu$ reduces the 
right hand side to $-\varepsilon_0\rho_1$ (where we assume unit normalised 
tangent vectors), so the projector equation \eqref{eqn:proj} is satisfied.

Again, each of the projectors~\eqref{eqn:typeCproj} reduces the rank of
$\varepsilon_0$ by half, giving eight, four and two $\aQ$ supercharges for the
plane, holomorphic curves in $\bR^4$ and $\bR^6$ respectively.

\paragraph[]{\hypertarget{tHsusy}{\tH}:}
Here we have arbitrary surfaces in $\bR^4$. Choosing $n$ according
to~\eqref{eqn:typeH}, for each tangent plane we can assign a linear 
combination of the projectors (see \eqref{eqn:eta})
\bal
\label{eqn:tHproj}
\tfrac{1}{2}(i \gamma_{12} \rho_3 +1)\,,\qquad
\tfrac{1}{2}(i \gamma_{31}\rho_2 + 1)\,,\qquad
\tfrac{1}{2}(i \gamma_{23}\rho_1 + 1)\,,\\
\tfrac{1}{2}(i \gamma_{34}\rho_3+1)\,,\qquad
\tfrac{1}{2}(i \gamma_{24}\rho_2+1)\,,\qquad
\tfrac{1}{2}(i \gamma_{14}\rho_1+1)\,.
\eal
In contrast to the \tR\ oscillators above, all these commute rather than
anticommute.

Not all the projectors are independent. For a generic surface in
$\bR^4$, we need to satisfy all of them.  Taking any projector of the
first line and its complement on the second line imposes $\varepsilon_0 =
-\varepsilon_0 \gamma_{1234}$, which is a restriction to the supersymmetries
antichiral with respect to 4d chirality. On these supersymmetries, all the
projectors of the second line and equal to those of the first, so we are left
with 3 independent conditions on the antichiral spinors. Hence generic surfaces
preserve a single $\aQ$.

For a Lagrangian submanifold of \tL\ one component of $n$ is zero (e.g. $n^3 =
0$). In that case we only need to take 2 columns of projectors out
of~\eqref{eqn:tHproj}, and this imposes again 2 conditions on antichiral
spinors, so leaving two $\aQ$.

Conversely if $n^3 = 1$ is constant (and all other components vanish), then the
projector equation imposes only the first column of~\eqref{eqn:tHproj}, and by
the discussion above it preserves four $\aQ$.
These are holomorphic curves, as in \tC\ (except for the choice of $n^3$ instead 
of $n^1$).

The \tN\ surfaces have the projectors on the first line of \eqref{eqn:tHproj}, which are all 
independent, so a generic surface of this type preserves two $\aQ$ supercharges. Note that 
the projectors on the second line are the same as \eqref{eqn:typeRproj} of \tR\
if we replace $x^6 \leftrightarrow x^4$ in that construction, also preserving
two $\aQ$s and manifesting another example of intersections between our
examples.

\paragraph[]{\hypertarget{cone}{Conical surfaces:}}
The discussion so far focused only on the Poincar\'e supercharges $\aQ$, but if the surfaces 
are conical with the tip at the origin, they also preserve conformal supercharges $\aS$ (for other locations 
of the tip, they would preserve other combinations of $\aQ$ and $\aS$). We can 
express the conical surfaces by $x^\mu= u \xi^\mu(v)$ where $\xi$ is a curve on an appropriate 
sphere.

The simplest example is the plane, where $\xi$ is a circle centred around the origin. 
In addition to the eight $\aQ$'s, it is annihilated by eight $\aS$'s. 
The only other examples of \tR\ are when the curve $x^\mu(u)$ is comprised of two rays meeting at any angle, 
which is a slight generalisation of the example mentioned in the introduction. The surface is then two 
half-planes glued along a line, so can be called \hypertarget{crease}{{\bf crease}}. The base of the cone  
$\xi^\mu$ is comprised of two half-circles on $S^2$, so two longitude lines.

To see the extended supersymmetry, specializing to a conical surface over $\xi^\mu$, 
\eqref{eqn:proj} becomes ($\dot\xi=d\xi/dv$)
\beq
\left[ \varepsilon_0 + \bar{\varepsilon}_1 (u\,\xi\cdot \bar{\gamma}) \right] 
\left[\im\varepsilon^{ab} \gamma_{\mu\nu}
\xi^\mu \dot \xi^\nu
+ n_I \rho_I \right] = 0\,.
\eeq
Of course $\xi^\rho\bar\gamma_\rho \gamma_{\mu\nu}\xi^\mu \dot \xi^\nu=\xi^\nu\bar\gamma_\nu$, 
but more importantly, we can also permute the first gamma matrix through, such that the equation 
becomes
\beq
\varepsilon_0 \left[\im\varepsilon^{ab} \gamma_{\mu\nu}\xi^\mu \dot \xi^\nu+ n_I \rho_I \right] 
+\bar{\varepsilon}_1
\left[-\im\varepsilon^{ab} \bar\gamma_{\mu\nu}\xi^\mu \dot \xi^\nu+ n_I \rho_I \right]  (u\,\xi\cdot \bar{\gamma})
= 0\,.
\label{eqn:coneproj}
\eeq
So the equations for $\varepsilon_1$ involve the complementary projectors to $\varepsilon_0$, 
and certainly the dimension of the space of solutions for the two are the same.

\paragraph{\bf Tori:}
Within the \tL\ surfaces in $\bR^4$, the simplest representatives are 
tori which are the product of circles in the $(x^1,x^2)$ and $(x^3,x^4)$ planes. In addition to the 
two $\aQ$ supercharges that any of the Lagrangian surfaces preserve, these also preserve a 
pair of $\aS$ generators. The derivation is similar to~\eqref{eqn:coneproj} and
uses that for these surfaces $x^2 = 1$.

\paragraph[]{\hypertarget{tSsusy}{\tS}:}
To understand this class, recall that it includes 1/2 BPS spheres, which are not annihilated by any one 
$\aQ$ or $\aS$, but by 16 linear combinations of the form $\aQ+\aS$. We can get this by a stereographic 
map of the plane, and likewise we can understand the surfaces foliated by arcs between the north and 
south pole by their stereographic projection to cones of \tN, which preserve two $\aQ$ and two $\aS$. 
So the analogs on $S^3$ preserve four independent combinations $\aQ+\aS$. As a special 
case, the conformal map of the \hyperlink{crease}{crease} give surfaces comprised of two 
hemispheres of different great $S^2$ and preserves eight combinations $\aQ + \aS$ 
(to visualise this, think of two longitudinal lines, or hemicircles, meeting at the north and south pole, 
and replace them with hemispheres).

The most general surface on $S^3$ preserves two supercharges. To see this,
write the projector equation~\eqref{eqn:proj} as
\beq
\label{eqn:Sproj1}
 \frac{1}{2}\varepsilon^{ab}\partial_a x^\mu \partial_b x^\nu
 \left[ \varepsilon_0 + \bar{\varepsilon}_1 (x
\cdot \bar{\gamma}) \right]\left[
\im\gamma_{\mu\nu}
+ \varepsilon_{\mu\nu\sigma I}x^\sigma \rho_I \right] = 0\,.
\eeq
We want this equation to hold for all orientations, meaning all $\mu,\nu$, so expanding and 
reorganising (assuming orthogonality) we find
\beq
\left[ \varepsilon_0 \im\gamma_{\mu\nu}
+ \bar{\varepsilon}_1 \bar{\gamma}^\sigma
\varepsilon_{\mu\nu\sigma I} \rho_I \right]
+\left[ \varepsilon_0 \varepsilon_{\mu\nu\sigma I}\rho_I
+\im \bar{\varepsilon}_1 \bar\gamma_{\mu\nu\sigma}
\right] x^\sigma = 0\,.
\eeq
Since $\varepsilon_0, \bar{\varepsilon}_1$ are constant the two terms in brackets
need to vanish separately, and in fact imposing one implies the other.

To count the number of independent constraints, note that referring to the
ansatz~\eqref{eqn:typeS} every tangent plane on $S^3$ is associated to a vector
$n$ on $S^3$. For the latter there is obviously four possible independent vectors,
so for a generic surface we impose four independent conditions on
$\varepsilon_0, \bar{\varepsilon}_1$. This leaves 2 preserved
supercharges.

\section{Anomaly}
\label{sec:anomaly}

Recall that surface operators have Weyl anomalies whose form is 
constrained by the Wess-Zumino consistency conditions to be a sum of conformal
invariants~\cite{deser:1993yx,Boulanger:2007st,Schwimmer:2008yh,Drukker:2020dcz}.
They can be expressed as the integral of the density
\beq
{\cal A}_{\Sigma}
  = \frac{1}{4\pi} \left[
  a_1{\cal R}^{\Sigma}
  + a_2 \left( H^2 + 4 \tr{P} \right)
  + b \tr{W}
  + c \cAnomaly \right].
  \label{eqn:anomaly0}
\eeq
The first invariant is $\RicciScalar^\Sigma$, which is the Ricci scalar of the
induced metric $h_{ab}$. The second piece includes $H^\mu$, the mean curvature
vector and $\tr{P} = h^{ab} P_{ab}$, the trace of the pullback of the Schouten
tensor. $W$ is the pullback of the Weyl tensor and $(\partial n)^2 = h^{ab}
\partial_a n^I \partial_b n^I$ is the norm of the variation of $n$ over the
surface. 

We work on flat euclidean space so $P = W = 0$. It is useful in the following to
have an explicit expression for the other invariants, and they can be
written nicely in terms of the second fundamental form $\IIFundForm$ as
\beq
  \IIFundForm^\mu_{ab} = \partial_a \partial_b x^\nu
  (\delta_\nu^\mu - h^{cd} \partial_c x_\nu \partial_d x^\mu)\,,\qquad
  H^\mu = h^{ab} \IIFundForm_{ab}^\mu\,,\qquad
  \RicciScalar^\Sigma = \IIFundForm_{ab}^\mu \IIFundForm_{cd}^\mu
  \varepsilon^{ac} \varepsilon^{bd}.
  \label{eqn:geom2ff}
\eeq
The term $(\delta_\nu^\mu - h^{cd} \partial_c x_\nu \partial_d x^\mu)$ is a
projector to the components normal to the surface.

Another simplification occurs in our case.
While the Wess-Zumino consistency condition allow the four 
independent coefficients $a_1$, $a_2$, $b$ and $c$ above, 
supersymmetry imposes $a_2=-c$ and $b=0$ for locally BPS surface operators
\cite{Bianchi:2019sxz,Drukker:2020atp}. The anomaly density then reduces to
\beq
  \mathcal{A}_\Sigma = \frac{1}{4 \pi} \left[
  a_1 \RicciScalar^{\Sigma} + c \left( \partial n^2 - H^2\right)
  \right].
  \label{eqn:anomaly}
\eeq
The constants $a_1$ and $c$ are observables that depend on the theory 
(the choice of ADE algebra underlying the $\cN = (2,0)$ theory), and the 
representation of the surface operator (at least for the $A_N$ theories they are 
classified by representations of that algebra). According to
\cite{Gentle:2015jma,Rodgers:2018mvq,Jensen:2018rxu,Estes:2018tnu,Chalabi:2020iie,Wang:2020xkc}
they are given by
\beq
  a_1 = \frac{1}{2} (\Lambda,\Lambda)\,, \qquad
  c = C_2(\Lambda) - a_1\,,
  \label{eqn:a1c}
\eeq
where $\Lambda$ is the highest weight defining the representation and
$C_2(\Lambda)$ is the quadratic Casimir of the representation. For the fundamental 
representation of $\su(N)$ this is
\beq
  a_1^{(N)} = \frac{1}{2} -\frac{1}{2N}\,, \qquad
  c^{(N)} = N-\frac12-\frac{1}{2N}\,,
  \label{eqn:a1cfund}
\eeq
but we leave them as $a_1$ and $c$ in the following.

The above expressions give a precise expression for the anomaly of all the surfaces 
studied in this paper, requiring just the evaluation of the integrals of the 
expressions in \eqref{eqn:anomaly}. 
But as in all our examples the vector $n$ depends in varying ways on $x^\mu$, we 
can evaluate $(\partial n)^2$ relying on the geometry as studied in 
Section~\ref{sec:geom} and simplify the expression for the anomaly even further.

\paragraph[]{\tR:}  For BPS surfaces satisfying~\eqref{eqn:typeR}, the anomaly
vanishes since $(\partial n)^2 = H^2$ and the Ricci scalar is zero.
In the abelian theory one can make a stronger statement, and using the field theory
propagators given in~\cite{henningson:1999xi,gus03,gustavsson:2004gj,Drukker:2020dcz}
one can check that the expectation value also vanishes identically.  It is
tempting to conjecture that this is true in general.

\paragraph[]{\tC:}  In this case $n^I$ is a constant, so $(\partial n)^2$
vanishes. $H^2$ also vanishes, which can be seen by a direct computation, or by
noting that the K\"ahler form $\omega$~\eqref{eqn:kahlerformr6} act as a
calibration for these surfaces, so that any holomorphic curve is a minimal
surface. The only contribution to the anomaly then comes from the Ricci scalar,
so it is proportional to the genus of the curve
\beq
  \int_\Sigma \mathcal{A}^\bC_\Sigma \vol_\Sigma =
  a_1 \chi(\Sigma)\,.
  \label{eqn:typeCanomaly}
\eeq
Note that for non-compact surfaces the genus is defined by this 
integral, and may not match that expected for the analogous curve in $\bC\bP^3$.

\paragraph[]{\tH:}
Here the anomaly is more interesting. Using~\eqref{eqn:typeH} and the definition
of the geometric invariants~\eqref{eqn:geom2ff}, we can write
\begin{align}
  (\partial n)^2 - H^2 = \eta_{\mu\nu}^I \eta^I_{\rho \sigma}
  (\IIFundForm_{ab}^\mu \partial_e x^\nu)
  (\IIFundForm_{cd}^\rho \partial_f x^\sigma)
  \left(\varepsilon^{be} \varepsilon^{df} h^{ac} - \frac{1}{2} h^{ab} h^{cd}
  h^{ef}\right)\,.
\end{align}
The 't~Hooft symbol \eqref{eqn:eta} satisfy
\begin{align}
  \eta_{\mu\nu}^I \eta^I_{\rho \sigma} =
  \delta_{\mu\rho} \delta_{\nu\sigma} - \delta_{\mu\sigma} \delta_{\nu\rho}
  + \varepsilon_{\mu\nu\rho\sigma}.
\end{align}
If we focus on the $\delta$ contractions, only the first is nonzero. We identify
$\delta_{\nu\sigma} \partial_e x^\nu \partial_f x^\sigma = h_{ef}$, and this
reduces the last term to $-\varepsilon^{ac}\varepsilon^{bd}$, so we immediately
recognize $-\RicciScalar^\Sigma$ from~\eqref{eqn:geom2ff}.

The contraction involving $\varepsilon_{\mu\nu\rho\sigma}$ gives us 
the Hodge dual of the 2-form $(\IIFundForm_{cd}^\rho \partial_f x^\sigma)
\diff x^\rho \diff x^\sigma$.  This interchanges the tangent and normal
space, and one can verify that it calculates the curvature of the
normal bundle, which has a coordinate expression
\beq
  \RicciScalar^\Sigma_\perp = -\frac{1}{2} \varepsilon_{\mu\nu\rho\sigma}
  (\IIFundForm^\mu_{ab} \IIFundForm^\nu_{cd}
  h^{bc} \varepsilon^{ad})
  \varepsilon^{ef}\partial_e x^\rho \partial_f x^\sigma \,.
  \label{eqn:selfinter}
\eeq
Its integral, the Euler class of the normal bundle, is also equal in 4d to the self-intersection number 
(denoted $[\Sigma] \cdot[\Sigma]$), see e.g.~\cite{bott2013differential}.

The term proportional to $c$ in the anomaly density is then
\beq
(\partial n)^2 - H^2 = -\RicciScalar^\Sigma + \RicciScalar^\Sigma_\perp.
\eeq
Including the $a_1$ anomaly and integrating over the surface, we obtain a topological quantity
\beq
  \int_\Sigma \mathcal{A}^\bH_\Sigma \vol_\Sigma =
  (a_1 - c) \chi(\Sigma) + c [\Sigma] \cdot [\Sigma].
  \label{eqn:typeHanomaly}
\eeq
This has a natural interpretation in terms of the Gauss map. Recall the image of the Gauss
map is two $S^2$'s and $n^I$ is in one of these
spheres. The topological invariant calculated by integrating
$-\RicciScalar^\Sigma + \RicciScalar^\perp$ is (-2 times) the degree of this map $\Sigma \to
S^2$.

It is interesting to compare this result with our discussion for holomorphic
surfaces in $\bC^2$. As argued around~\eqref{eqn:typeCanomaly}, for holomorphic
surfaces $(\partial n)^2 - H^2 = 0$. Indeed in this case the Gauss map gives 
a constant $n$ and certainly the degree of the map vanishes, and indeed 
the self-intersection is equal to the euler characteristic. The same is true
also for Lagrangian submanifolds, as the image is on the equator of $S^2$, so
again the degree vanishes.

\paragraph[]{\tS:}
Here we simply expand $(\partial n)^2$ using~\eqref{eqn:typeS} to obtain
\beq
  \left( \partial n \right)^2 =
  \partial_a \partial_b x^\mu \left( \delta^{\mu\nu} - h^{cd} \partial_c x^\mu
  \partial_d x^\nu \right) \partial_e \partial_f x^\nu h^{ae} h^{bf} - 2\,.
\eeq
The factor 2 seems surprising at first, as it leads to a term in the anomaly 
proportional to the area of the surface $\Sigma$. This in fact is not in contradiction 
to conformal invariance, as this construction has a dimensionful parameter, the
radius of the sphere (which we set to one).  As for the first term, we can identify 
the second fundamental form from~\eqref{eqn:geom2ff} to rewrite this as
\beq
  \left( \partial n \right)^2 - H^2 =
  - \RicciScalar^\Sigma - 2\,.
\eeq
The anomaly is then
\beq
\label{eqn:typeSanomaly}
  \int \mathcal{A}^S_\Sigma \vol_\Sigma =
  \left( a_1 - c \right) \chi(\Sigma) -
  c \frac{\vol(\Sigma)}{2 \pi}\,.
\eeq

For infinitesimal surfaces on $S^2$ we should recover the anomaly of~\tN. In that limit the area 
of the surface vanishes  
giving~\eqref{eqn:typeHanomaly}, since in 3d the self-intersection
number~\eqref{eqn:selfinter} is always zero.

\subsection[Conical singularities]{\hyperlink{cone}{Conical singularities}}
\label{sec:cone}

Surfaces with conical singularities have two sources of anomalies. First there
are the usual anomalies discussed above but whose density may have a singular
contribution at the apex of the cone.  Second, the tips are distinguished
points, so we may find new anomalies when integrating finite quantities over a
scale-invariant cone. In fact, for non-BPS surfaces these two sources of
divergences can conflate to $\log^2\epsilon$ divergences~\cite{klebanov:2012yf,
myers:2012vs,bianchi:2015liz,dorn:2016bkd,Drukker:2020dcz}.  For BPS surfaces
they split into two independent contributions. We discuss them in turn.

We start with a cone over a curve $\cK\subset S^2$ (parametrised by $\xi^\mu$)
of \tN, which is a subclass of \tH. The self-intersection
number~\eqref{eqn:selfinter} vanishes in three dimensions, so the anomaly
\eqref{eqn:typeHanomaly} is determined solely by the Euler characteristic. The
Ricci scalar vanishes along the cone, but diverges at the tip. It can be
regularised by smoothing the tip at distances $u\leq\epsilon$ to form the
regularised surface $S_\cK$. Using the Gauss-Bonnet theorem
\beq
  \frac{1}{4\pi} \int_{S_\cK} \mathcal{R}^\Sigma \vol_\Sigma
  = \chi(S_\cK) - \frac{1}{2 \pi} \int_{\cK} \kappa_g \vol_\cK.
\eeq
Note that we defined here the anomaly for the non-compact cone without a boundary 
term at infinity, which would be on the left hand side, so instead it is on the right. If we choose 
the disc topology for $S_\cK$, the first term gives $1$.

To evaluate the boundary contribution we can take our parametrisation $\xi^\mu$
to be unit speed, so that the geodesic curvature $\kappa_g$ is constant and
equal to $1/\epsilon$ when evaluated on the sphere of radius $\epsilon$. The
line integral then calculates the arc length of the curve $\cK$, which is
$\epsilon L_\cK$ ($L_\cK$ being the arc length on a unit sphere).

Therefore the anomaly is
\beq
\int_\Sigma {\cal A}_\Sigma \vol_\Sigma =
\left(1- \frac{L_\cK}{2\pi}\right) (a_1 - c) \,.
\eeq

The story for general cones of \tH\ is more interesting, since the curve $\cK$
can have the topology of a nontrivial knot on $S^3$.  In the case of the unknot,
we could choose the disc topology for our regularisation, but more generally
this is a Seifert surface (see e.g.  \cite{adams1994knot}). Given a knot there
are many different possible such surfaces of different topology. This is easy to
see already in the case of the disk to which we could add any number of handles,
which would modify the anomaly above to 
\beq
\left(1-2n- \frac{L_\cK}{2\pi}\right) (a_1 - c) \,,
\qquad n\in{\mathbb N}\,.
  \label{eqn:typeHanomalycones}
\eeq
For a general knot there would be Seifert surfaces of lowest genus, whose 
Euler characteristic would replace the 1 in this equation, and other regularisations would 
still allow for the integer $n$.  A slight generalisation to this result is to
allow for multiple curves ${\cK_i}$ defining a link. There is again a Seifert
surface of lowest genus whose boundary is the link, and in addition one needs to
replace $L_\cK$ by $\sum_i L_{\cK_i}$.

The discussion thus far focused on regular surface anomalies. They are localized at the apex of the 
cone and can be calculated by choosing an appropriate regularisation of the
conical singularity. The second type of divergences arising from cones come from 
terms that would be finite upon regularisation. Consider a cone with no anomaly
density (apart from the apex) and restrict to an open subset $u\in(u_{min},u_{max})$.
While the contribution to the expectation value is finite for any such open
subset, it would naturally scale like $\log(u_{min}/u_{max})$ and, after taking the
limit $u_{min} \to \epsilon$, contribute a new term to the anomaly.

An example of such surfaces are the cones of \tH. They are related to BPS Wilson
loops in 5d, see the discussion in Section~\ref{sec:Wilson} below.

We can consider also surfaces that are not globally conical, but have conical
singularities.  Examples are the surfaces foliated by hemicircles of \tS,
conformal to \tN\ cones. Those have the topology of the sphere, so we expect
that the anomaly does not require special regularization.

\subsection{Anomaless surfaces}

Surfaces that are not anomalous can have finite expectation values, which one may hope to be 
able to calculate exactly, as is done for BPS line operators in 3d and 4d (as well as 
other observables).

In the examples above we found that those of \tR\ are indeed anomaless, 
though we also expect their expectation value to vanish identically.
This is certainly true in the free abelian theory, where using the propagators
of~\cite{Drukker:2020dcz} and the ansatz~\eqref{eqn:typeR} one can check that
the propagator for the 2-form $B_{\mu\nu}$ exactly cancels the scalar $\Phi^I$ propagator.

Both the anomaly of \tC\ \eqref{eqn:typeCanomaly} and 
\tH\ \eqref{eqn:typeHanomaly} are purely topological, so any surface with
vanishing $\chi(\Sigma)$ and self-intersection number is anomaless. This
includes all (topological) tori of \tH, with examples in \tL\ and \tN.

The formula for the anomaly of \tH\ \hyperlink{cone}{\bf cones}, \eqref{eqn:typeHanomalycones} 
vanishes only for $n=0$ and $L_{\cal K}=2\pi$. This means it is a cone over the 
unknot with zero deficit angle and includes the crease 
of arbitrary opening angle (and in particular the plane). 
These surfaces cannot have finite expectation values, as they are non-compact, 
Note that by conformally transforming to a compact manifold like 
the sphere (or the crease generalisation of 
it discussed above equation \eqref{eqn:Sproj1}) changes the topology, so those 
examples are in fact anomalous.

Finally, we can look for cases where the surfaces are anomaless only for particular values 
of $a_1$ and $c$ \eqref{eqn:a1c}. For example, if we restrict to surfaces in the fundamental 
representation, $c^{(N)}=(2N+1)a_1^{(N)}$ \eqref{eqn:a1cfund}. Now looking 
at the \tS\ anomaly \eqref{eqn:typeSanomaly}, this vanishes for
\beq
\frac{\vol(\Sigma)}{4\pi}=-\chi(\Sigma)\frac{N}{2N+1}\,.
\eeq
For example for $\chi=-2$, or genus 2 surfaces, the resulting area would 
be close to that of a maximal $S^2$. For higher genus, the surfaces would 
have to be even larger.

For \tH\ surfaces the analogous equation \eqref{eqn:typeHanomaly} vanished for
\beq
[\Sigma] \cdot [\Sigma]=\chi(\Sigma)\frac{2N}{2N+1}\,.
\eeq
The simplest possible solution would require Euler characteristic $\chi=-4N-2$ and 
self-intersection number $[\Sigma] \cdot [\Sigma]=-4N$. We do not know whether 
there are solutions to these conditions. One can of course also go to other 
representations using the expressions in \eqref{eqn:a1c}.

The expressions in \eqref{eqn:a1c} do not extend to the abelian theory, where 
$a_1 = c=1/2$. In this case any surface of \tH\ with vanishing self-intersection number 
is anomaless. In the large $N$ limit $a_1\ll c$, so the classical supergravity 
calculation \cite{graham:1999pm, Drukker:2020dcz} is only sensitive to the 
$c$ anomaly. In this limit all \tH\ surfaces of equal Euler character and self-intersection 
number as well as all \tC\ surfaces are effectively anomaless.

\section{Dimensional reduction}
\label{sec:reduction}
The inspiration for this project comes from the rich history of BPS Wilson loops in $\cN=4$ SYM in 
4d \cite{erickson:2000af, drukker:2000rr, pestun:2007rz, zarembo:2002an,drukker:2007qr}. 
It is therefore natural to examine whether any of the families of surfaces uncovered here 
arise as the dimensional uplift of BPS loops in 4d or 5d. Likewise we would like to address the 
relation to BPS surfaces in lower dimensions.

\subsection{Wilson and 't~Hooft loops}
\label{sec:Wilson}

The classes of BPS loops in $\cN=4$ SYM have been completely classified \cite{Dymarsky:2009si} and 
it is easy to check that they are indeed related to \tR\ and conical \tH\ surfaces, 
as follows.

To get line operators in lower dimensions we need to wrap the surface operator on a compact 
direction, so they need to be homogeneous in one direction. The simplest example is 
\tR\ surfaces when the theory is compactified along the $x^6$ direction. This gives line 
operators in 5d and upon further reduction, lines in 4d. As the 4d gauge coupling is related 
to the modulus of the compactification torus, depending on which cycle of the torus is 
$x^6$, we end up with either Wilson, 't~Hooft or dyonic operator. The Wilson loops are clearly 
of the class identified by Zarembo in \cite{zarembo:2002an}, the the 't~Hooft loops are their S-duals 
\cite{Kapustin:2006pk}.
Note that \tH\ surfaces extended along say $x^4$ are identical to \tR\ ones in $\bR^3$ with 
$x^6$ replaced by $x^4$.

Part of the reason we expect \tR\ surfaces to have vanishing expectation value in 6d is that 
this is true for the loops in 4d \cite{Guralnik:2003di, Guralnik:2004yc, Dymarsky:2006ve}.

The other example arises from \hyperlink{cone}{\bf conical surfaces} extended along the 
radial direction. We can act with a conformal transformation that maps $\bR^6\to \bR\times S^5$ 
by taking the log of the radial coordinate. Conical surfaces are mapped to surfaces extending 
along $\bR$ and wrapping a curve in $S^5$. Compatifying, we end up with 5d Yang-Mills on 
$S^5$, and the surface operators are mapped to Wilson loops in that theory.

Recall that conical surfaces may have two types of divergences. Anomalies arising from singular 
curvature at the apex are now lost, as we changed the topology of the surface to a cylinder. 
The second type of divergences are easier to understand in this cylinder picture, as a finite 
expectation value for the 5d Wilson loops uplifts to a finite result for the compactified 
surfaces. The expression is necessarily extensive in the compactification radius and in the 
uncompactified limit, which is conformal to $\bR^6$ studied in this paper, diverges. This 
is the source of the second type of cone anomalies alluded to in Section~\ref{sec:cone}.

Wilson loops in 5d following the 4d ansatz 
in \cite{drukker:2007qr} and their uplift to surface operators in 6d were studied 
in \cite{mezei:2018url}. That construction is also based on the chiral $\eta$ symbol and indeed maps to 
our \tH\ cones. Using localization, one is able to evaluate the Wilson loops explicitly 
and the answer is proportional to that of Wilson loops in 3d topological Chern-Simons theory 
\cite{Witten:1988hf}. The proportionality constant can indeed be identified with the 
compactification radius.

Carrying this over to our setup, we find that for conical \tH\ surfaces, in addition to the 
usual anomaly, which as explained in Section~\ref{sec:cone} is related to the 
arc-length of the base of the cone and the genus of an appropriate Seifert surface, 
there is another logarithmic divergence proportional to the topological Chern-Simons 
Wilson loop. Note that this localised divergence does not follow the structure of 
\eqref{eqn:anomaly0} and in particular the dependence on $N$ and the representation 
is not captured by \eqref{eqn:a1c}.

The conical surfaces of \tH\ realise the loops of \cite{mezei:2018url} in 5d, but not the original 
ones in 4d \cite{drukker:2007qr}. The reason is that the preserved supersymmetries 
mix Poincar\'e and conformal supercharges (similar to \tS\ surfaces, see \eqref{eqn:Sproj1}). 
As conformal invariance in 4d and 6d are not directly related, the uplift to surfaces on 
$S^3\times \bR\subset\bR^6$ is not BPS but $S^3\times S^1\subset\bR^4\times T^2$ is 
(or as mentioned above, on $S^3\times \bR\subset S^5\times\bR$).

Compactifying the theory on other Riemann surfaces leads to class-$\cal S$ theories and 
BPS line operators in those theories \cite{Drukker:2009tz} originate from BPS surfaces in 6d. 
The supercharges surviving the required topological twist are those satisfying 
\eqref{eqn:topotwist}. They  are compatible with a surface operator along an appropriate cycle on the 
Riemann surface \cite{Drukker:2009tz}, an arbitrary curve in $\bR^2\subset\bR^4$ and the adaptation 
or \eqref{eqn:typeR} to couple to $n^4$ and $n^5$. This give the class 
of mutually-BPS line operators in $\cN=2$ theories \cite{Kapustin:2006hi}, 
which follow the same 
ansatz as \cite{zarembo:2002an}, or our \tR, but restricted to a curve in $\bR^2$.

\subsection{Surfaces in 4d}
\label{sec:4dsurf}
The other way to dimensionally reduce a surface is if it is at a fixed point along a compact direction. 
In that case we would get a surface operator in the lower dimensional theory. This can give surface 
operators in any dimension, but let us focus on 4d, either $\cN=4$ theory or
theories of class-$\cal S$. Theories with $\cN=1$ supersymmetry in 4d also have 
BPS surface operators, but we will not touch upon those. 
As the construction of surfaces of \tH\ and \tS\ is inherently four dimensional, we expect them to be 
realised in 4d as well.

A first comment is that when restricting to $\bR^4$ both \tC\ and \tR\ become subclasses of 
\tH, and \tR\ is essentially the same as \tN. 
This was discussed already below equations \eqref{eqn:eta} and~\eqref{eqn:tHproj} respectively. 
So we need only discuss \tH\ (and its subclasses: \tC, \tR, \tL\ and \tN) and separately \tS.

A second comment is that within the AGT correspondence \cite{alday:2009aq} it is very natural to distinguish 
the origin of a surface operator in 4d as arising from a codimension-4 operator in 6d, 
or from a a codimension-2 defect wrapping the Riemann surface. While our discussion 
is explicitly relevant only for the former case, it is only based on supersymmetry analysis, 
so is very likely to carry over, as to any other operator involving the coupling 
of an appropriate 2d superconformal field theory to the 4d theory.

Let us separate the discussion to theories of 
class-$\cal S$ and $\cN=4$ SYM.

\subsubsection*{$\cN=2$ theories:}
There is an immense literature on surface operators in 4d class-$\cal S$ theories 
focusing mostly on operators breaking $\su(2,2|2)\to\su(2|1)^2$ 
(see, e.g. \cite{Gaiotto:2009fs,Gomis:2014eya,Gukov:2014gja,LeFloch:2020uop}).
This is indeed 
the symmetry preserved by the 1/2 BPS surface operator of the 6d theory appropriately 
compactified to $\cN=2$ in 4d. In 2d language, these are $\cN=(2,2)$ defects.

To make the relation between 6d and 4d explicit, it involves
replacing the $x^5$, $x^6$ directions with a compact Riemann surface. 
In order to preserve supersymmetry away from
flat space, one performs a topological twist which preserves the supercharges
satisfying
\beq
  \varepsilon (\gamma_{56} + \rho_{45}) = 0\,.
  \label{eqn:topotwist}
\eeq
The resulting supersymmetries are the $\aQ$ and $\bar{\aQ}$
of the 4d theory. It is a simple check that~\eqref{eqn:topotwist}
is automatically satisfied if we impose all the projectors~\eqref{eqn:tHproj}.

Viewing things from the 4d point of view, the equations of BPS surface operators 
are very similar to \eqref{eqn:proj}, 
with only three $\rho$'s and the four dimensional Killing spinors. 
As mentioned above, all but \tS\ are subclasses of \tH, so we analyse the latter first in turn.

\paragraph[]{\tH:}
For the most general surface of this type $n^I\in S^2$, and we need to solve 
all the equations of~\eqref{eqn:tHproj}. The projectors in the first column give 
the 4d chirality projector $\gamma_{1234}$, eliminating the $\bar\aQ$ supercharges. 
This is not surprising as the \tH\ ansatz \eqref{eqn:typeH} is chiral. 
In 4d the three projectors on the first line of~\eqref{eqn:tHproj} are not independent, as 
the gamma matrices of $\su(2)_R$ symmetry satisfy $\rho_1\rho_2\rho_3=-i$. 
We find that the generic surface of this type preserves a single $\aQ$

\paragraph[]{\tR:}
Here we have surfaces extended along $x^4$ and an arbitrary surface in $\bR^3$ and generically 
$n^I\in S^2$. As already mentioned, these surfaces are now a 
subclass of \tH\ and one should impose the projector equations on the second line 
of~\eqref{eqn:tHproj}. Given that $\varepsilon_0$ 
is comprised of one 4-component chiral and one 4-component antichiral 
spinors, the two independent equations in the second line reduce to 
one $\aQ$ and one $\bar\aQ$.

\paragraph[]{\tN:}
This case is equivalent to \tR\ in four dimensions, except that we need to impose the 
conditions on the \textit{first} line of \eqref{eqn:tHproj}.

\paragraph[]{\tL:}
In this case we need to impose the equations associated to two of the columns 
in~\eqref{eqn:tHproj}. As mentioned above, these automatically imply that the third 
column is also satisfied, so there is no supersymmetry enhancement for 
Lagrangian surfaces.

\paragraph[]{\tC:}
For holomorphic surfaces which are not a plane, we need to impose both equations in one 
of the columns in~\eqref{eqn:tHproj}, so we find that the surfaces preserve 
a pair of $\aQ$ supercharges. Indeed holomorphic surface
operators in $\cN=4$ SYM were constructed in \cite{Koh:2008kt} and the analysis 
here suggests that this should extend to any class-$\cal S$ theory.

\paragraph[]{\tS:}
In this case the ansatz \eqref{eqn:typeS} requires $n^I\in S^3$, so cannot be fully realised 
for $\cN=2$ theories. Recall though that surfaces of \tN\ are conformal to ``banana'' 
surfaces on $S^3$ and for those $n^4=0$. We can implement this conformal 
transformation for $\cN=2$ theories in 4d as well, so they too preserve a pair 
of superchages: one $\aQ+\aS$ and one $\bar\aQ+\bar\aS$.

\subsubsection*{$\cN=4$ SYM:}
The dimensional reduction on a torus to $\cN=4$ theories does not break any supersymmetries, 
thus guaranteeing that our analysis carries over. The most studied surface operators 
are those of Gukov and Witten \cite{Gukov:2006jk}, and though they arise from codimension-2 
defects in six dimensions, the 1/2 BPS plane preserves the same superalgebra as arising 
from 6d surface operators, $\su(2|2)^2$, or in two dimension language $\cN=(4,4)$.

The projector equations in 4d take a slightly different form from \eqref{eqn:proj}, as is clear, 
since the R-symmetry breaking is $\mathfrak{so}(6)\to\mathfrak{so}(4)$, which requires two $\rho$ matrices. 
Also, the structure of the Killing spinors is slightly different. Yet, the dimensional reduction 
guarantees that solutions to all our classes exist. The number of preserved supercharges 
is also slightly different from 6d, and is double the number listed for the appropriate subclasses 
of \tH\ for $\cN=2$ theories above.

\paragraph[]{\tS:}
In this case the we can realise the full ansatz \eqref{eqn:typeS} 
and the supersymmetry analysis in Section~\ref{sec:susy} is indeed 
consistent in 4d. The most general surface preserves two supercharges.

We should mention that this supersymmetry analysis does not provide a construction 
of these operators, but suggests which symmetries to try to realise. It would 
be interesting to extend the constructions of 
\cite{Gukov:2006jk,Gukov:2008sn} as defect operators and those of 
\cite{Gaiotto:2009fs, Gomis:2014eya} as coupled 2d-4d systems to arbitrary 
SUSY preserving geometries, generalising the \tH\ case of \cite{Koh:2008kt}.

\section{Holography}
\label{sec:holography}

The large $N$ limit of the $\cN=(2,0)$ theory has a holographic representation in terms 
of M-theory on $AdS_7\times S^4$~\cite{maldacena:1997re}. Therein, the surface operators are 
described by M2-branes ending along the surface at the boundary of space 
\cite{maldacena:1998im}. Globally BPS surface operators should be described 
by M2-brane embeddings which realise the same supersymmetries. 

Here we take the first few steps to realise the different classes of BPS
surfaces. The simplest examples of M2-branes embeddings are those
whose boundary is a plane or a sphere, they were presented respectively in
\cite{maldacena:1998im} and \cite{berenstein:1998ij}.

We take again inspiration from the holographic realisation of the large $N$ 
limit of BPS Wilson loops in $\cN=4$ SYM. 
For each of the construction of \cite{zarembo:2002an, drukker:2007qr}, the BPS 
conditions on the field theory could be extended to an almost complex structure on a 
subspace of $AdS_5\times S^5$. In both cases the dual classical string surfaces are 
pseudo-holomorphic with respect to the almost complex structure. Furthermore, 
the action of the string can be evaluated by the pullback of an appropriate 
form in the bulk, which manifests as a (generalised) calibration. In the following 
we present some analogous structures for the four main classes of BPS surface 
operators. We refer to the original papers~\cite{Dymarsky:2006ve,drukker:2007qr}, 
for more details on the string constructions.

Our starting point is the metric on $AdS_7 \times S^4$,
\beq
ds^2=\frac{y}{L}\,\diff x^\mu\diff x^\mu
+\left(\frac{L}{y}\right)^2\diff y^I\diff y^I\,,
\label{eqn:metric}
\eeq
where $y \equiv |y^I|$.
To get the regular Poincar\'e patch metric redefine $y=4L^3/z^2$ and parametrise 
$y^I/y$ as angular coordinates on $S^4$. Note that here $L$ is the radius of $S^4$ and the 
curvature radius of $AdS_7$ is $2L$.
The background also has a 4-form field strength 
proportional to the volume form on $S^4$.

The Killing spinors in this coordinate system can be constructed from the same
pair of constant spinors $\varepsilon_0$ and $\bar{\varepsilon}_1$ of Section~\ref{sec:susy}. 
Using the 11-dimensional curved space $\Gamma$-matrices satisfying
\begin{align}
  \left\{ \Gamma_\mu, \Gamma_\nu \right\} = \frac{2 y}{L} \delta_{\mu\nu}\,,
  \qquad
  \left\{ \Gamma_\mu, \Gamma_I \right\} = 0\,, \qquad
  \left\{ \Gamma_I, \Gamma_J \right\} = 2 \left( \frac{L}{y} \right)^2
  \delta_{IJ}\,,
\end{align}
they are given by \cite{Claus:1998yw}
\beq
  \varepsilon =
  \left(\frac{y}{L}\right)^{1/4} \varepsilon_0
  +\left( \frac{L}{y} \right)^{1/4} \bar{\varepsilon}_1 \left(x^\mu
  \Gamma_\mu-2y^I\Gamma_I\right)\,.
\eeq
In the limit of $y\to\infty$, the last term drops out and we recover the
conformal Killing spinors of flat space appearing in~\eqref{eqn:proj},
$(y/L)^{1/4}(\varepsilon_0+ \bar{\varepsilon}_1 x^\mu \gamma_\mu)$. 
This enables a very direct map between the BPS conditions that we found 
for the different types of surfaces in Section~\ref{sec:susy} to the bulk.

In the following we use the notation $X^M$ for coordinates of $AdS_7 \times S^4$ and index $m =
1,2,3$ for the M2-brane. $G_{MN}$ is the metric~\eqref{eqn:metric} and $g_{mn}$
is the induced metric on the M2-brane.

The analog of the projector equation in supergravity is a consequence of the 
$\kappa$-invariance of the M2-brane action~\cite{Bergshoeff:1987cm}. It reads
\beq
  -\frac{i}{6} \varepsilon \Gamma_{MNP} \partial_m X^M \partial_n X^N \partial_p X^P
  \varepsilon^{mnp} = \varepsilon\,,
\label{eqn:kappa}
\eeq
where $\varepsilon^{mnp}$ is the Levi-Civita tensor density and includes
$1/\sqrt{g}$. On the boundary this equation reduces to~\eqref{eqn:proj}, so the
supercharges preserved by the M2-brane are constructed from the same
$\varepsilon_0$ and $\bar{\varepsilon}_1$ as in field theory.

Given a preserved supercharge represented by $\varepsilon$, we can construct a 
3-form~\cite{mezei:2018url}
\beq
  \phi = -i \,\frac{\varepsilon \Gamma_{MNP} \varepsilon^\dagger}{\varepsilon \varepsilon^\dagger}\,
  \diff X^M \wedge \diff X^N \wedge \diff X^P\,.
  \label{eqn:phidef}
\eeq
By construction, the pullback of this 3-form to an M2-brane satisfying~\eqref{eqn:kappa} 
is its volume form.  $\phi$ then serves a role analogous to the almost complex
structure in string theory.

If $\phi$ is closed $\diff \phi = 0$, then the form is a
calibration~\cite{harvey1982calibrated,Dymarsky:2006ve,joyce2007riemannian}; its integral is the 
same on all 3-surfaces of the same homology class and is equal to the volume 
of the minimal surface, so the classical M2-brane action.

If $\phi$ is exact then the action comes only from the boundary
at large $y$ and is simply proportional to $y\vol(\Sigma)$. This 
divergent term is removed by the Legendre transform
of~\cite{drukker:1999zq,Rodgers:2018mvq,Drukker:2020dcz} or equivalently by
renormalisation~\cite{Bianchi:2001kw}, and the expectation value of the surface
operator vanishes.

In each of our main examples we construct the appropriate $\phi$ and 
comment on its properties. It is particularly useful when one can write 
an equation of the form
\beq
  \partial_m X^M =
  \frac{1}{2} g_{ml} \varepsilon^{lnp} G^{ML} \phi_{LNP}
  \partial_n X^N \partial_p X^P\,.
  \label{eqn:g2structurePDE}
\eeq
This replaces the equations of motion with a set of first order nonlinear equations 
and is the generalisation of the pseudo-holomorphicity condition in string theory. 
The exact form of this equation varies between the examples presented below.

\paragraph[]{\tR:}
Using $\varepsilon$ with $\varepsilon_0$ annihilated by all the $a_J$ 
oscillators in \eqref{eqn:ladderalgebra}, equation \eqref{eqn:phidef} gives 
the 3-form
\beq
  \phi^\bR =
  -\diff x^6\wedge  \omega^\bR\,,
  \qquad
  \omega^\bR=\sum_{I=1}^5\left(\diff x^I\wedge dy^I \right)\,.
  \label{eqn:typeR3form}
\eeq
$\phi^\bR= - \diff(x^6 \omega^\bR)$ is clearly exact, so the expectation value of the 
\tR\ surface operators vanishes at large $N$.

By translation symmetry, the 3-surface should extend in the $x^6$ direction 
and identifying $\sigma^3=x^6$, we expect the remaining coordinates to 
satisfy the pseudo-holomorphicity condition
\beq
  \partial_m X^M = g_{ml} \varepsilon^{ln3} G^{ML} \omega^\bR_{LN} \partial_n X^N\,.
\label{eqn:typeRformPDE}
\eeq

\paragraph[]{\tC:}
In this case there are two preserved supercharges, so~\eqref{eqn:phidef} gives two choices of 3-form. The 
components shared by both are
\beq
  \varphi^\bC =
  (\diff x^1\wedge\diff x^2 + \diff x^3 \wedge \diff x^4 + \diff x^5\wedge\diff x^6) \wedge \diff y^1\,.
\label{eqn:typeC3form0}
\eeq
This is obvious uplift of the the K\"ahler form $\omega^\bC$ of
$\bR^6$~\eqref{eqn:kahlerformr6} to $AdS_7$: 
$\varphi^\bC=\omega^\bC\wedge\diff y^1$. It is natural to augment it to
\beq
  \phi^\bC =
  (\diff x^{12} + \diff x^{34} + \diff x^{56}) \wedge \diff y^1
  +\left( \frac{y}{L} \right)^{3/2} (\diff x^{136}+\diff x^{145}+\diff
  x^{235}-\diff x^{246})\,,
  \label{eqn:typeC3form}
\eeq
which has the form of a $G_2$-structure (the shorthand notation is $\diff
x^{\mu\nu\dots}= \diff x^\mu \wedge \diff x^\nu \wedge \dots$).

In this case the first order equation \eqref{eqn:g2structurePDE} is indeed satisfied. This 
is a consequence of properties of $G_2$-structures as described below, 
around~\eqref{eqn:a^2}, but this equation is 
also satisfied with the restricted version of $\varphi^\bC$ with just the three components 
\eqref{eqn:typeC3form0}.
That 3-form is closed, which means that minimal volumes are parallel to it.
Choosing a gauge where $X^7=y^1=\sigma^3$, this is the requirement that $\partial_3 x^\mu =
0$, so the pullback of $\varphi^\bC$ is
\beq
\varphi^\bC_{\mu\nu7}
\partial_m x^\mu \partial_n x^\nu \,\diff\sigma^m\wedge\diff \sigma^n\wedge\diff \sigma^3\,.
\label{eqn:typeCformPDE}
\eeq
This is the pullback of the K\"ahler form $\omega^\bC$ to the surface at fixed $y^1$ times $\diff\sigma^3$.
We can immediately integrate over $\sigma^3$ and, using~\eqref{eqn:holomorphic},
simply find a linear divergence times the area element of the 2-surface
$\Sigma$. The expectation value of the surface operator therefore vanishes, which is also
clear, since this version of the 3-form is exact. This is consistent with
\eqref{eqn:typeCanomaly}, as this calculation captures only ${\cal O}(N)$ terms
in the large $N$ limit~\cite{graham:1999pm,Drukker:2020dcz} and the anomaly
coefficient $a_1^{(N)}\to1/2$ \eqref{eqn:a1cfund}.

\paragraph[]{\tH:}

The natural 3-form extending \eqref{eqn:tHproj} to $AdS_5\times S^2$ is
\begin{align}
  \label{eqn:typeH3form}
  \phi^\bH &=
  \eta_{\mu\nu}^I\,\diff x^\mu \wedge \diff x^\nu \wedge dy^I
  - \frac{L^3}{y^3}\diff y^{123}\\&
  = (\diff x^{12} + \diff x^{34}) \wedge \diff y^3+(\diff x^{31} + \diff
  x^{24}) \wedge \diff y^2 + (\diff x^{23} + \diff x^{14}) \wedge \diff y^1
  -\frac{L^3}{y^3}\diff y^{123}\,.
\nonumber
\end{align}
This is indeed the result of~\eqref{eqn:phidef} with the supercharge preserving 
all \tH\ surfaces \eqref{eqn:tHproj}. This form is closed, so defines a 
calibration, and as in the \tC\ above~\eqref{eqn:typeC3form}, this is a 
$G_2$-structure~\cite{bonan1966varietes,bryant1989,Hitchin:2000jd,joyce2000compact}
(see also~\cite{Karigiannis_2020} for a gentle introduction).

Let us prove this implies that the first order equation \eqref{eqn:g2structurePDE} 
is satisfied. Following~\cite{Dymarsky:2006ve}, consider the quantity
\beq
  a_m^M = \partial_m X^M - \frac{1}{2} g_{ml} \varepsilon^{lnp} G^{ML} \phi_{LNP}
  \partial_n X^N \partial_p X^P\,.
\eeq
Squaring this we get
\bal
\label{eqn:a^2}
  g^{mn} a_m^M a_n^M G_{MN} &=
  g^{mn} \partial_m X^M \partial_n X^N G_{MN}
  - \varepsilon^{mnp} \partial_m X^M \partial_n X^N \partial_p X^P \phi_{MNP} \\
  &\quad+ \frac{1}{2} g^{mn} g^{pq} \partial_m X^M \partial_n X^N \partial_p X^P
  \partial_q X^Q
  G^{RS} \phi_{RMP} \phi_{SNQ}\,.
\eal
The first term gives the trace of the induced metric and the second 
using~\eqref{eqn:phidef}, the contraction of two Levi-Civita tensors. Their sum
is $-3$. The terms of the second line can be evaluated using properties of the
3-form $\phi$. In particular $G_2$-structures satisfy
\begin{align}
G^{RS} \phi_{RMP} \phi_{SNQ}
=  G_{MN} G_{PQ} - G_{MQ} G_{PN} - (* \phi)_{MPNQ}\,.
  \label{eqn:g2phirel}
\end{align}
The asymmetric term does not contribute and the contractions with the first two reduce 
everything to traces of the induced metric and the second line evaluates to 3, showing that 
$a_m^M=0$, thus proving~\eqref{eqn:g2structurePDE}.

Surfaces analogous to \hyperlink{cone}{\bf cones} of \tH\ were studied in \cite{mezei:2018url} 
(see the discussion in Section~\ref{sec:Wilson}), including their holographic 
realisation. There it was shown that the M2-branes are calibrated with respect to a 3-form, which should 
be the same as \eqref{eqn:typeH3form} in a different coordinate system. A reduction 
of~\eqref{eqn:g2structurePDE} to conical surfaces (which looks similar to 
\eqref{eqn:typeRformPDE}) was also derived there.

\paragraph[]{\tS:}

These surfaces are restricted to $S^3 \subset \bR^4$ and $n^I\in S^3$, so the M2-brane 
duals are located within an $AdS_4 \times S^3\subset AdS_7\times S^4$. Using the 
metric~\eqref{eqn:metric}, we take the eight coordinates $x^I$ and $y^I$ with 
$I=1,2,3,4$, set $x^5=x^6=y^5=0$ and impose the further constraint
\beq
  |x|^2 + \frac{4L^3}{y} = 1\,.
\label{eqn:restriction}
\eeq

Using the two supercharges that preserve the \tS\ surfaces \eqref{eqn:Sproj1} gives according 
to \eqref{eqn:phidef} two 3-forms. The terms appearing in both are
\beq
  \phi^S= \frac{1}{6} \varepsilon_{IJKL} \left[
    \diff x^I \diff x^J \left( 3 x^K \diff y^L-2y^K \diff x^L \right)
    - \left( \frac{L}{y} \right)^3 \diff y^I \diff y^J
    \left(x^K \diff y^L  -6 y^K \diff x^L \right)
  \right]\,.
\label{eqn:phiS}
\eeq
To simplify this, we define the complex vielbeins $\zeta^I$
\beq
\zeta^I= \sqrt{\frac{y}{L}}\,\diff x^I+i\frac{L}{y}\diff y^I\,,
\qquad
\bar\zeta^I= \sqrt{\frac{y}{L}}\,\diff x^I-i\frac{L}{y}\diff y^I\,.
\eeq
They are null with respect to the metric \eqref{eqn:metric}, and
$\frac{i}{2}\sum_i \zeta^I\wedge\bar\zeta^I$ defines an almost complex structure
on $AdS_5\times S^3$.  Taking $\Xi$ to be the unit vector normal to
\eqref{eqn:restriction}, its contraction with $\zeta^I$ gives
\beq
\Xi\cdot\zeta^I
=\sqrt{\frac{L}{y}}\left(x^J\partial_{x^J}-2 y^J\partial_{y^J}\right)\cdot\zeta^I
=x^I-2i\left(\frac{L}{y}\right)^{3/2}y^I\,,
\eeq
and the complex conjugate for $\Xi\cdot\bar\zeta^I$.

Now taking the (anti)holomorphic 4-forms on $AdS_5\times S^3$
\beq
\Omega=\zeta^1\wedge\zeta^2\wedge\zeta^3\wedge\zeta^4\,,
\qquad
\bar\Omega=\bar\zeta^1\wedge\bar\zeta^2\wedge\bar\zeta^3\wedge\bar\zeta^4\,,
\eeq
$\phi^S$ is the projection of their difference onto the hypersurface \eqref{eqn:restriction}
\bal
\phi^S  
&=\frac{1}{2i}\,\Xi\cdot\left(\Omega-\bar\Omega\right)
=\frac{\im}{12}\varepsilon_{IJKL}
\left(\zeta^I\wedge\zeta^J\wedge \zeta^K \left(\Xi\cdot\zeta^L\right)
-\bar\zeta^I\wedge\bar\zeta^J\wedge\bar\zeta^K
\left(\Xi\cdot\bar\zeta^L\right)\right)\,.
\label{eqn:phiScomplex}
\eal
This representation makes it manifest that though we write $\phi$ in terms of eight coordinates, 
it is contained within (the cotangent to) the $AdS_4\times S^3$ subspace defined by
\eqref{eqn:restriction}. Perhaps it is consistent with some notion of
generalised calibrations as in \cite{Gutowski:1999iu,Gutowski:1999tu,
drukker:2007qr}.

\paragraph{Minimal surfaces:}
The discussion above allowed us to reduce the BPS equations to first order equations 
for \tR, \tC\ and \tH. The structure of 3-form for \tS\ in~\eqref{eqn:phiS} is more complicated, 
and it is not clear what should be the form of~\eqref{eqn:g2structurePDE} in this case. 

The equations for \tR\ and \tC\ can be written as pseudo-holomorphicity equations 
\eqref{eqn:typeRformPDE}, \eqref{eqn:typeCformPDE} and in the latter 
case the solutions are simply $\Sigma\times\bR_+$ where $\Sigma$ is 
the original surface and 
$y^1\in\bR_+$. For \tC\ the equation can also be written in a form identical 
to~\eqref{eqn:g2structurePDE} and this is also the form of the equation for \tH\ surfaces.

We leave it to future work to find explicit solutions to those equations.

\section{Discussion}
\label{sec:discussion}
This paper is meant to reveal some cracks in the steep granite face of the
$\cN=(2,0)$ theory. They are unlikely to offer an easy path to the summit, but
the anchors we placed should provide a good starting route for further
exploration. At the very least, they promise to be a fun playground for those
interested in this mysterious theory.

In the absence of a fully satisfying Lagrangian (or other traditional) formulation, 
we relied here on symmetry, geometry and algebra, the knowledge that 
these theories have planar surface operator observables and the large symmetry 
that they preserve. It is a very natural (and mild) assumption that such operators 
can have arbitrary geometries and an associated vector field $n$ representing 
the local R-symmetry breaking. Indeed this assumption can easily be checked 
and realised in both the abelian theory and the holographic realisation.

This philosophy leads to trying to find geometries and vector fields for which 
the projector equation \eqref{eqn:proj} has global solutions. To our surprise 
we found a very intricate structure of solutions to this set of equations 
which we split into several main classes, subclasses and examples preserving 
different numbers (and types) of supersymmetries, see Tables~\ref{tab:types} 
and~\ref{tab:examples}.

The most obvious question is whether the examples we found are exhaustive. 
In one sense the answer is clearly no: acting with a global symmetry gives more 
examples and we chose to focus on single representatives of those. While 
changing $x^6$ to $x^5$ in \tR\ or the three components of $n^I$ that are turned on by 
the \tH\ ansatz are rather trivial, acting with conformal transformations are less so. 
For example, they map $\bR^4$ of \tH\ to $S^4$ and for surfaces that extend to 
infinity, they change the topology, so can affect the anomaly, or in the absence 
of an anomaly the finite expectation value. A simple illustration of this is the relation between the
non-anomalous plane and its anomalous counterpart, the sphere.
Likewise for \tC\ surfaces, their 
compact versions could also be interesting.

Another generalisation is to allow $n$ to be complex, while keeping $(n)^2 = 1$ 
(which is required in order to satisfy \eqref{eqn:proj}).
In this case the analytic continuation to Lorentzian signature would lead to 
non-unitary operators, but the analog for those in the case of Wilson 
loops have been instrumental for localisation \cite{pestun:2007rz}, the 
ladder limit \cite{Correa:2012nk} and the related fishnet theories 
\cite{Zamolodchikov:1980mb, Mamroud:2017uyz, Gromov:2017cja}.

Complexifying $n$ is somewhat akin to smearing chiral operators over the surface, 
but in fact one can consider combinations of surface operators and local chiral 
operators. Those can either be operators constrained to the surface such that they do not 
exist in its absence, or be away from it. The former case is studied extensively for 
maximally symmetric setups as part of the defect CFT programme 
\cite{Cardy:1991tv,McAvity:1995zd,Liendo:2012hy, Gaiotto:2013nva, Billo:2016cpy,
Bianchi:2019sxz,Drukker:2020atp}. 
This could be extended to general surfaces and indeed a natural approach to try
to prove that \tR\ surfaces have trivial expectation value is to view all
surfaces of this type as deformations of one another by the insertion of
appropriate combinations of geometric and R-symmetry-changing local operators.
In the latter case the question is which combination of surfaces and local
operators are mutually BPS along the lines of
\cite{Semenoff:2001xp,Drukker:2009sf, Giombi:2012ep, Beem:2013sza, 
Beem:2014kka, Giombi:2018qox}.

Our approach here was to try auspicious ans\"atze and were fortunate to uncover 
some examples, but it should be possible to formalise the 
problem and treat it systematically. This could lead to a classification of BPS 
surface operators, as was done for Wilson loops in 4d \cite{Dymarsky:2009si}.

Another generalisation would be to study surfaces when the theory itself is on 
curved space. We touch on that when relating to surface operators in class-$\cal S$ 
theories in Section~\ref{sec:4dsurf}. One can likewise reduce the theory to 
3d \cite{Dimofte:2011ju} and 2d \cite{Benini:2013cda, Gadde:2013sca}, 
and one can look for other geometries it can be placed on and find compatible 
surface operators.

Lower dimensional line operator versions of \tR\ and conical \tH\ Wilson loops have 
been previously described in \cite{zarembo:2002an, Kapustin:2006pk, drukker:2007qr, mezei:2018url}
as were 4d surface operators similar to \tC\ in \cite{Koh:2008kt}. It would be interesting to realise the surface
operators of \tH\ and \tS\ there as well.

In four dimensional field theories entangling surfaces are also two dimensional and 
have been studied extensively in recent 
years~\cite{solodukhin:2008dh,myers:2012vs,bianchi:2015liz,Jensen:2018rxu}. 
They share many properties with surface operators, in particular the anomaly 
structure. Even more relevant for our discussion here is the supersymmetric 
R\'enyi entropy \cite{Nishioka:2013haa, Nishioka:2016guu}, 
and it would be interesting to see whether more results on entropy can 
be gleaned from entangling surfaces inspired by our different classes.

In the cases where we can define the $(2,0)$ theory, namely the abelian theory 
and the large $N$ holographic realisation, we can proceed to try to evaluate 
expectation values of surface operators explicitly. The anomaly has already 
been reproduced in both cases in \cite{graham:1999pm,henningson:1999xi,gustavsson:2004gj,Drukker:2020dcz}. 
To go beyond that, one can try to 
calculate finite expectation values for anomaless surfaces in the abelian theory 
employing the free-field propagators \cite{Drukker:2020dcz}. 
In the holographic setup it would be interesting to utilise the first order equations 
in Section~\ref{sec:holography} to find explicit examples, at least for highly 
symmetric surfaces. In addition to providing finite expectation values for 
anomaless surfaces, it would allow to calculate the correlation function with local 
operators as in \cite{berenstein:1998ij}.

One may hope to be able to realise the BPS surface operators within some old and new 
approaches to the $\cN=(2,0)$ theory
\cite{Aharony:1997an, ArkaniHamed:2001ie, Chu:2013hja,Lambert:2020scy,
Lambert:2020zdc,Alday:2020tgi}.
Many of those formulations address the presence of planar surfaces 
and it would be interesting to look for the richer spectrum of BPS observables found here.

Even limited to the abelian theory, we can use it to get clues to the nonabelian
theory. Following~\cite{Drukker:2007yx} we could look for an effective theory
calculating the BPS surface operators. An example is the proposal
of~\cite{mezei:2018url} that the (compactified) cones of \tH\ are calculated by a
3d Chern-Simons theory. Attempting this for \tS, we can write down an effective 
propagator which includes the contribution from the 2-form $B$ and the scalars $\Phi^I$ 
that couples to $n^I$ in \eqref{eqn:typeS} \cite{Drukker:2020dcz} to get an effective 2-form propagator
\beq
  \vev{B^{\mu \nu}(x) B_{\rho \sigma}(y)}
  = \frac{1}{4 \pi^2 \norm{x-y}^2} \left[
    \frac{1}{2} \delta_{\mu\nu}^{\rho\sigma} -
    \frac{4 \left( x-y \right)_{[\mu} \left( x-y \right)^{[\rho}
  \delta_{\nu]}^{\sigma]}}{\norm{x-y}^2} \right].
  \label{eqn:Propagator2form}
\eeq
This behaves somewhat like the propagator of a 2-form in 4d, and can also be dualised 
to a compact scalar. Integrating this over the surface easily reproduces the \tS\ anomaly 
in Section~\ref{sec:anomaly} including the area term~\eqref{eqn:typeSanomaly}. Finding an
appropriate nonabelian generalisation could help in guessing an expression for the 
surface operators a generic $(2,0)$ theory. 

Finally, the $\cN = (2,0)$ theory also contains co-dimension 2 defects, i.e. observables defined 
over 4-manifolds \cite{Witten:1997sc, Howe:1997hx, Bergshoeff:1997bh, gaiotto:2009we, Gaiotto:2009gz}
(see also \cite{Zhou:2015kaj}). 
It would be interesting to study their supersymmetric embeddings
in 6d, the corresponding anomalies and the compatibility with the surface
operators presented here.

\subsection*{Acknowledgments}
It is a pleasure to thanks Dima Panov and Simon Salamon for explaining to us a lot 
of the geometry underlying these constructions 
and to Malte Probst for continued discussion, related collaborations, and just for being an awesome guy.

ND's research is supported by the Science Technology \& Facilities council under the grants 
ST/T000759/1 and ST/P000258/1. 
MT acknowledges the support of the Natural Sciences and Engineering Research Council 
of Canada (NSERC).  
Cette recherche a \'et\'e financ\'ee par le Conseil de recherches en sciences naturelles 
et en g\'enie du Canada (CRSNG).

\bibliographystyle{utphys2}
\bibliography{ref}

\end{document}